\documentclass[twocolumn,aps,prl,superscriptaddress,showpacs,floatfix]{revtex4}
\usepackage{amssymb}
\usepackage{amsmath,bm}
\usepackage[normalem]{ulem}
\usepackage[dvips]{color}
\usepackage{booktabs}
\usepackage{graphicx}
\renewcommand{\sout}{\bgroup \color[rgb]{1,0,0}\ULdepth=-.5ex \ULset}

\begin{document}

\title{Probing QCD critical fluctuations from light nuclei production in relativistic heavy-ion collisions}

\author{Kai-Jia Sun\footnote{%
sunkaijia@sjtu.edu.cn}}
\affiliation{School of Physics and Astronomy and Shanghai Key Laboratory for
Particle Physics and Cosmology, Shanghai Jiao Tong University, Shanghai 200240, China}
\author{Lie-Wen Chen\footnote{%
Corresponding author: lwchen$@$sjtu.edu.cn}}
\affiliation{School of Physics and Astronomy and Shanghai Key Laboratory for
Particle Physics and Cosmology, Shanghai Jiao Tong University, Shanghai 200240, China}
\author{Che Ming Ko\footnote{%
ko@comp.tamu.edu}}
\affiliation{Cyclotron Institute and Department of Physics and Astronomy
Texas A\&M University, College Station, Texas 77843, USA}
\author{Zhangbu Xu\footnote{%
xzb@bnl.gov}}
\affiliation{Brookhaven National Laboratory, Upton, New York 11973, USA}
\affiliation{School of Physics \& Key Laboratory of Particle Physics and Particle Irradiation (MOE), Shandong University, Jinan, Shandong 250100, China}

\date{\today}

\begin{abstract}
Based on the coalescence model for light nuclei production, we show that the yield ratio $
\mathcal{O}_\text{p-d-t} = N_{^3\text{H}} N_p / N_\text{d}^2$ of $p$, d, and $^3$H in heavy-ion collisions is sensitive to the neutron relative density fluctuation $\Delta n= \langle (\delta
n)^2\rangle/\langle n\rangle^2$ at kinetic freeze-out. From recent experimental data in central
Pb+Pb collisions at $\sqrt{s_{NN}}=6.3$~GeV, $7.6$~GeV, $8.8$~GeV, $12.3$~GeV and $17.3$~GeV measured by the NA49 Collaboration at the CERN Super Proton Synchrotron (SPS), we find a possible non-monotonic behavior of $\Delta n$ as a function of the collision energy with a peak at $\sqrt{s_{NN}}=8.8$~GeV, indicating that the density fluctuations become the largest in collisions at this energy. With the known chemical freeze-out conditions determined from the statistical model fit to experimental data, we obtain a chemical freeze-out temperature of $\sim 144~$MeV and baryon chemical potential of $\sim 385~$MeV at this collision energy, which are close to the critical endpoint in the QCD phase diagram predicted by various theoretical studies.  Our results thus suggest the potential usefulness of the yield ratio of light nuclei in relativistic heavy-ion collisions as a direct probe of the large density fluctuations associated with the QCD critical phenomena.
\end{abstract}
\pacs{25.75.-q, 25.75.Dw}
\maketitle

Understanding the properties of strongly interacting matter under extreme conditions, particularly the phase transition between the quark-gluon plasma (QGP) and the hadronic matter, is a topic of great current interest~\cite{Shu09,Fuk11,Mun16,Pas17}. Results from lattice quantum chromodynamics (LQCD) calculations~\cite{Aok06,Bha14,For09,Fod04,Ale11,For14} and effective model studies~\cite{Asa89,Ste98,Ber99,Hat03,Ste04,Asa06} have indicated that the QGP to hadronic matter transition is likely first-order phase transition if the system has a large baryon chemical potential but changes to a crossover if its baryon chemical potential is small. This suggests the existence of a critical endpoint (CEP), where the first-order phase transition ends, in the temperature versus baryon chemical potential ($T,\mu_B$) plane of QCD phase diagram. To search for the CEP and locate its position in the QCD phase diagram, experiments have been carried out through the Beam Energy Scan (BES) program and will also be performed at the future Facility for Antiproton and Ion Research (FAIR) in Germany.

It has been argued that the enhanced long-wavelength fluctuations near the CEP can lead to singularities in all thermodynamic observables~\cite{Ste98}. The resulting event-by-event fluctuations of conserved quantities in relativistic heavy-ion collisions have thus been extensively studied both theoretically and experimentally. For example, the energy dependence of the fourth-order fluctuation~($\kappa\sigma^2$) of net-proton distribution measured in the BES program by the STAR Collaboration is found to exhibit the largest deviation from unity in Au+Au collisions at $\sqrt{s_{NN}}=7.7$ GeV~\cite{Luo15}. Also, owing to the different features between a first-order phase transition and a rapid crossover, one expects a non-monotonic behavior in the collision energy and centrality dependence of certain properties of the produced matter in heavy-ion collisions as it approaches the CEP, such as the ratio of its shear viscosity to entropy density~\cite{Cse06,Lac07}, expansion speed~\cite{Hun95,Ris96} and the slope of direct flow of light cluster~\cite{Bat16,Bas16}. Furthermore, a non-monotonic excitation function for the Gaussian emission source radii difference ($R_\text{out}^2-R_\text{side}^2$) extracted from two-pion interferometry measurements~\cite{Pra84,Cha95,Wie96} in Au+Au~($\sqrt{s_{NN}}$=7.7-200~GeV) and Pb+Pb~($\sqrt{s_{NN}}=2.76~$TeV) collisions has recently been observed with a maximum value located around $\sqrt{s_{NN}}=40~$GeV~\cite{Lac15}.

In analogy to the phenomenon of critical opalescence observed in the liquid-gas phase transition~\cite{And69,Ber09}, the matter created in relativistic heavy-ion collisions could develop large baryon density fluctuations when its evolution trajectory in the ($T,\mu_B$) plane of QCD phase diagram passes across the first-order phase transition line, especially when it is close to the CEP.  When the evolution trajectory approaches the CEP, the correlation length increases drastically, and the density fluctuation enhances accordingly and reaches its maximum value at the CEP. Studies based on both the hydrodynamic approach~\cite{Stei12,Stei14,Hero14} and the microscopic transport model~\cite{Li16} indeed show that the spinodal instability due to the first-order phase transition between the QGP and hadronic matter can induce large baryon density fluctuations. In the case that such density fluctuations can survive final-state interactions during the hadronic evolution of heavy-ion collisions, there should exist strong fluctuations in the nucleon density and thus significant inhomogeneity in the spatial distribution of nucleons at kinetic freeze-out. The baryon density fluctuations is, however, expected to be negligible if the QGP to hadronic matter transition is a crossover. Therefore, the nucleon density fluctuations at kinetic freeze-out in relativistic heavy-ion collisions may provide a unique probe to the critical endpoint in the QCD phase diagram.

In this Letter, we show for the first time that the relative density fluctuation of neutrons~($\Delta n= \langle (\delta n)^2\rangle/\langle n\rangle^2$) at kinetic freeze-out in relativistic heavy-ion collisions can be encoded in the yield ratio of light nuclei, namely, $\mathcal{O}_\text{p-d-t} = N_{^3\text{H}} N_p / N_\text{d}^2$. Our result thus has the advantage of directly measuring the density fluctuation instead of using the number fluctuation to infer the density fluctuation as having been done so far. Our study is based on the coalescence model for light nuclei production~\cite{But61,Sat81,Cse86,Dov91,ChenLW03,Gre03,Fri03,LWChen06,Fri08,Zhu15,KJSUN15,KJSUN17,Yin17}. In this model, the probability for the production of a nucleus depends on the nucleon many-body correlations and is thus affected by the fluctuations in the nucleon number or density. From analyzing the very recent data on the proton ($p$), deuteron (d) and triton (t or $^3$H) yields in Pb+Pb collisions at SPS energies measured by the NA49 Collaboration~\cite{Ant16}, we have observed a possible peak of $\Delta n$ in Pb+Pb collisions at $\sqrt{s_{NN}}= 8.8$~GeV. This result has further allowed us to estimate that the temperature and baryon chemical potential at which the CEP is located in the QCD phase diagram are $T^{\rm CEP} \sim 144$ MeV and $\mu_B^{\rm CEP} \sim 385$ MeV.

We start by briefly introducing the newly derived analytical coalescence formula COAL-SH~\cite{KJSUN17} for cluster production in relativistic heavy-ion collisions. In COAL-SH, the yield $N_c$ (per unit rapidity) of a cluster at midrapidity and consisting of $A$ constituent particles from the hadronic matter at kinetic freeze-out or emission source of effective temperature $T_{\rm eff}$ (including the effect of transversal flow), volume $V$, and number $N_i$ of the $i$-th constituent with mass $m_i$ reads~\cite{KJSUN17}
\begin{eqnarray}\label{Eq1}
N_c&=&g_\text{rel}g_\text{size}g_c M^{3/2}\bigg[\prod_{i=1}^{A} \frac{N_i}{m_i^{3/2}} \bigg]\nonumber\\
&\times&\prod_{i=1}^{A-1}\frac{\left(4\pi/\omega\right)^{3/2}}{Vx(1+x^2)} \left(\frac{x^2}{1+x^2}\right)^{l_i} G(l_i,x).
\end{eqnarray}
In the above, $M=\sum_{i=1}^{A}m_i$ is the rest mass of the cluster, $l_i$ is the orbital angular momentum associated with the $i$-th relative coordinate, $\omega$ is the oscillator frequency of the cluster's internal wave function and is inversely proportional to $M r_\text{rms}^2$ with $r_\text{rms}$ being the root-mean-square (RMS) radius of the cluster, and $G(l,x) = \sum_{k=0}^{l}\frac{l!}{k!(l-k)!} \frac{1}{(2k+1)x^{2k}}$ with $x=(2T_{\rm eff}/\omega)^{1/2}$ is the suppression factor due to the orbital angular momentum on the coalescence probability~\cite{Cho11}. In addition, $g_c =(2S+1)/(\prod_{i=1}^{A}(2s_i+1))$ is the coalescence factor for constituents of spin $s_i$ to form a cluster of spin $S$, $g_\text{rel}$ is the relativistic correction to the effective volume in momentum space, and $g_\text{size}$ is the correction due to the finite size of produced cluster.

In central Pb+Pb collisions considered here, $V$ is much larger than the sizes of light nuclei, and we thus set $g_\text{size} = 1$. We also set $g_\text{rel} = 1$ because the masses of nucleons and light nuclei are much larger than the value of $T_{\rm eff}$. For light nuclei included in the present study, all the constituent nucleons are in $s$-state ($l=0$), and we thus have $G(l,x)=1$. From Eq.~(\ref{Eq1}), the yields of d and $^3$H are then simply given by
\begin{eqnarray}
N_{\rm d} &=& g_{\rm d}\frac{(m_n+m_p)^{3/2}}{m_p^{3/2} m_n^{3/2}} \frac{N_p N_n}{V}\frac{(4\pi/\omega_{\rm d})^{3/2}}{x_{\rm d}(1+x_{\rm d}^2)},\label{Eq2}\\
N_{^3\text{H}} &=& g_{^3{\rm H}}\frac{(2m_n+m_p)^{3/2}}{m_p^{3/2} m_n^{3}} \frac{N_p N_n^2}{V^2}\frac{(4\pi/\omega_{\rm ^3H})^{3}}{x_{\rm ^3H}^2(1+x_{\rm ^3H}^2)^2}, \label{Eq3}
\end{eqnarray}
where $N_p$ ($N_n$) is the number of protons (neutrons) in the emission source, the coalescence factor
is $g_{\rm d}=3/4$ for d and $g_{^3{\rm H}}=1/4$ for $^3$H, and we denote $x_{\rm d}=(2T_{\rm eff}/\omega_{\rm d})^{1/2}$ and $x_{\rm ^3H}=(2T_{\rm eff}/\omega_{\rm ^3H})^{1/2}$ with the oscillator frequency $\omega_{\rm d} = 8.1$ MeV for d and $\omega_{\rm ^3H}= 13.4$ MeV for $^3$H obtained from their respective RMS radii $r_\text{rms,d} = 1.96$ fm and $r_\text{rms,$^3$H} = 1.76$ fm~\cite{Rop09}. The effective temperature $T_{\rm eff}$ at the kinetic freeze-out in relativistic heavy-ion collision is typically about $200$~MeV and is thus much larger than the oscillator frequencies $\omega_{\rm d}$ and $\omega_{\rm ^3H}$. Neglecting neutron and proton mass difference ($m_p=m_n=m_0$) and noting $x_{\rm d}, x_{\rm ^3H}\gg 1$, we then have
\begin{eqnarray}
N_{\rm d} &=& \frac{3}{2^{1/2}}\left(\frac{2\pi}{m_0T_{\rm eff}}\right)^{3/2}\frac{N_p N_n}{V},\label{Eq4}  \\
N_{^3\text{H}} &=& \frac{3^{3/2}}{4}\left(\frac{2\pi}{m_0T_{\rm eff}}\right)^{3}\frac{N_p N^2_n}{V^2}. \label{Eq5}
\end{eqnarray}
Although the coalescence formula COAL-SH is derived by assuming the Bjorken boost invariance~\cite{Bjo83} for the emission source, Eqs.~(\ref{Eq4}) and (\ref{Eq5}) turn out to be also valid for an isotropically expanding fireball. This is not surprising as only the effective temperature, volume, and proton and nucleon numbers appear in these equations. Also, the above equations are consistent with the predictions from the thermal (statistical) model~\cite{Bra07,And11,Cle11,Ste12} if $p$, $n$, d and $^3$H are assumed to be in thermal and chemical equilibrium and the binding energies of d and $^3$H are neglected.

\begin{table*}
\caption{Yields~($dN/dy$ at midrapidity) of $p$, d, $^3$He and $^3$H as well as the yield ratio $^3$H/$^3$He measured in Pb+Pb collisions at SPS energies~\cite{Ant16} together with the derived yield ratio $\mathcal{O}_{\text{p-d-t}}$. The units for $E$ and $\sqrt{s_{NN}}$ are AGeV and GeV, respectively.}
\begin{tabular}{c|c|c|c|c|c|c|c|c}
        \hline \hline
           $E$&$\sqrt{s_{NN}}$  & centrality&  $p$ & d & $^3\text{He}$ & $^3\text{H}/^3\text{He}$ &$^3$H & $\mathcal{O}_{\text{p-d-t}}$ \\
         \hline
         20 & 6.3& $0-7\%$&    46.1$\pm$2.1 & 2.094$\pm$0.168& $3.58(\pm0.43)\times 10^{-2}$&1.22$\pm$0.10& $4.37(\pm0.64)\times 10^{-2}$& 0.459$\pm$0.014  \\
         30 & 7.6&$0-7\%$&    42.1$\pm$2.0 & 1.379$\pm$0.111& $1.89(\pm0.23)\times 10^{-2}$&1.18$\pm$0.11& $2.23(\pm0.34)\times 10^{-2}$& 0.494$\pm$0.020  \\
         40 &8.8&$0-7\%$&     41.3$\pm$1.1 & 1.065$\pm$0.086& $1.28(\pm0.15)\times 10^{-2}$&1.16$\pm$0.15& $1.48(\pm0.26)\times 10^{-2}$& 0.541$\pm$0.022 \\
         80 &12.3& $0-7\%$&    30.1$\pm$1.0 & 0.543$\pm$0.044& $3.90(\pm0.50)\times 10^{-3}$&1.15$\pm$0.19& $4.49(\pm0.94)\times 10^{-3}$& 0.458$\pm$0.038 	\\
        158 &17.3& $0-12\%$&   23.9$\pm$1.0 & 0.279$\pm$0.023& $1.50(\pm0.20)\times 10^{-3}$&1.05$\pm$0.15& $1.58(\pm0.31)\times 10^{-3}$& 0.484$\pm$0.037 \\
        \hline  \hline
\end{tabular}
\label{tab1}
\end{table*}

\begin{table*}
\caption{Collision energy dependence of neutron relative density fluctuation $\Delta n$ for $\alpha = $-0.2, -0.1, 0, 0.1 and 0.2. The units for $E$ and $\sqrt{s_{NN}}$ are AGeV and GeV, respectively.}
\begin{tabular}{c|c|c|c|c|c|c|c}
        \hline \hline
           $E$&$\sqrt{s_{NN}}$  & centrality& $\Delta n$~($\alpha = -0.2$)& $\Delta n$~($\alpha = -0.1$)& $\Delta n$~($\alpha = 0$)& $\Delta n$~($\alpha = 0.1$)& $\Delta n$~($\alpha = 0.2$) \\
         \hline
         20 & 6.3& $0-7\%$&   0.485$\pm$0.037&   0.526$\pm$0.039&   0.583$\pm$0.048&   0.669$\pm$0.064&   0.816$\pm$0.099 \\
         30 & 7.6&$0-7\%$&   0.566$\pm$0.044&   0.623$\pm$0.053&   0.704$\pm$0.068&   0.833$\pm$0.096&   1.093$\pm$0.177 \\
         40 &8.8&$0-7\%$&   0.667$\pm$0.046&   0.746$\pm$0.057&   0.864$\pm$0.076&   1.071$\pm$0.118&   1.620$\pm$0.322\\
         80 &12.3& $0-7\%$&  0.482$\pm$0.090&  0.523$\pm$0.106&  0.579$\pm$0.130&  0.662$\pm$0.171&  0.807$\pm$0.262	\\
        158 &17.3& $0-12\%$& 0.542$\pm$0.084& 0.594$\pm$0.101& 0.668$\pm$0.127& 0.782$\pm$0.175& 1.002$\pm$0.345\\
        \hline  \hline
\end{tabular}
\label{tab2}
\end{table*}

In obtaining Eqs.~(\ref{Eq4}) and (\ref{Eq5}), we have assumed that nucleons are uniformly distributed in space at kinetic freeze-out. To take into account density fluctuations of nucleons, we express the neutron and proton density in the emission source as
\begin{eqnarray}
n(\vec{r}) &=& \frac{1}{V}\int n(\vec{r})\text{d}\vec{r}+\delta n(\vec{r}) = \langle n\rangle+\delta n(\vec{r}), \label{Eq6}\\
n_p(\vec{r}) &=& \frac{1}{V}\int n_p(\vec{r})\text{d}\vec{r}+\delta n_p(\vec{r}) = \langle n_p\rangle+\delta n_p(\vec{r}), \label{Eq7} 
\end{eqnarray}
where $\langle \cdot\rangle$ denotes the average value over space and $\delta n(\vec{r})$~($\delta n_p(\vec{r})$) with $\langle\delta n\rangle=0$~($\langle\delta n_p\rangle=0$) denotes the fluctuation of neutron~(proton) density from its average value $\langle n\rangle$~($\langle n_p\rangle$).
We can then approximately rewrite Eqs.~(\ref{Eq4}) and (\ref{Eq5}) as
\begin{eqnarray}
N_{\rm d} &=& \frac{3}{2^{1/2}}\left(\frac{2\pi}{m_0T_\text{eff}}\right)^{3/2} \int \text{d}\vec{r} ~n(\vec{r})n_p(\vec{r})  \notag \\
&=& \frac{3}{2^{1/2}}\left(\frac{2\pi}{m_0T_\text{eff}}\right)^{3/2} ~(N_p\langle n\rangle + V\langle\delta n \delta n_p\rangle),  \label{Eq8}
\end{eqnarray}
and
\begin{eqnarray}
N_{\rm ^3H} &=& \frac{3^{3/2}}{4}\left(\frac{2\pi}{m_0T_\text{eff}}\right)^{3} \int \text{d}\vec{r} ~n(\vec{r})^2n_p(\vec{r})  \notag \\
&=& \frac{3^{3/2}}{4}\left(\frac{2\pi}{m_0T_\text{eff}}\right)^{3} \big[ (\langle n\rangle^2+\langle(\delta n)^2\rangle)N_p \notag \\ && +2V\langle n\rangle\langle\delta n \delta n_p\rangle+ V\langle(\delta n)^2\delta n_p\rangle\big]. \label{Eq9}
\end{eqnarray}

Assuming $\delta n_p(\vec{r})=c(\vec{r})\delta n(\vec{r})$, where the function $c({\vec{r}})$ can be positive or negative, we can then express the correlation between $\delta n(\vec{r})$ and $\delta n_p(\vec{r})$ as
\begin{eqnarray}
\langle\delta n \delta n_p\rangle &=& \frac{1}{V}\int \text{d} \vec{r}\delta n(\vec{r}) \delta n_p(\vec{r}) \notag \\
&=& \frac{1}{V}\int \text{d} \vec{r}~c(\vec{r})(\delta n(\vec{r}))^2.  \label{Eq9-1}
\end{eqnarray}
The above equation can also be written as
\begin{eqnarray}
\langle\delta n \delta n_p\rangle = \alpha \frac{\langle n_p\rangle}{\langle n\rangle}\langle (\delta n)^2\rangle , \label{Eq10}
\end{eqnarray}
with $\alpha$ being the correlation coefficient and $\frac{\langle n_p\rangle}{\langle n\rangle}$ accounting for the isospin asymmetry of the emission source. In the case that the neutron and proton density fluctuations are completely correlated, we then have $\alpha = 1$. By neglecting the term $\langle (\delta n)^2 \delta n_p\rangle$ in Eq.~(\ref{Eq9}), we can rewrite Eqs.~(\ref{Eq8}) and (\ref{Eq9}) as
\begin{eqnarray}
N_{\rm d} &=&\frac{3}{2^{1/2}}\left(\frac{2\pi}{m_0T_\text{eff}}\right)^{3/2} ~N_p\langle n\rangle (1+\alpha \Delta n),   \label{Eq11} \\
N_{\rm ^3H} &=&  \frac{3^{3/2}}{4}\left(\frac{2\pi}{m_0T_\text{eff}}\right)^{3}
N_p\langle n\rangle^2 [1+(1+2\alpha)\Delta n], \notag \\ \label{Eq12}
\end{eqnarray}
where $\Delta n=\langle (\delta n)^2\rangle /\langle n\rangle^2$ is a dimensionless quantity that characterizes the relative density fluctuation of neutrons.

Besides depending on $\Delta n$, both d and $^3$H yields also depend on $T_\text{eff}$, $N_p$ and $\langle n\rangle$. The density fluctuation in the emission source can be probed from the following yield ratio:
\begin{eqnarray}
\mathcal{O}_\text{p-d-t} = \frac{N_{^3\text{H}}N_p}{N_\text{d}^2}&=&g\frac{1+(1+2\alpha)\Delta n}{(1+\alpha \Delta n)^2}, \label{Eq13}
\end{eqnarray}
with $g=4/9\times(3/4)^{3/2}\approx 0.29$. The $\mathcal{O_\text{p-d-t}}$ is constructed in such a way that many effects, such as those due to $T_\text{eff}$, $N_p$, $\langle n\rangle$, volume and isospin asymmetry of the emission source, cancel out. Experimentally, one can thus extract $\Delta n$ in relativistic heavy-ion collisions by measuring the yield ratio $\mathcal{O}_\text{p-d-t}$. When $\alpha\Delta n$ is much smaller than unity, the correction from $\alpha$ in Eq.~(\ref{Eq13}) is second-order, and $\mathcal{O}_\text{p-d-t}$ can be approximated as
\begin{eqnarray}
\mathcal{O}_\text{p-d-t} \approx g(1+\Delta n). \label{Eq14}
\end{eqnarray}
In this case, $\mathcal{O}_\text{p-d-t}$ has a very simple linear dependence on $\Delta n$. We would like to point out that one may also choose other light nuclei such as $^3$He and $^4$He to extract the nucleon density fluctuation at kinetic freeze-out. In these cases, however, information on the isospin at freeze-out is needed and also the higher-order density fluctuations may be involved. For example, the yields of $^3$He and $^4$He are given, respectively, by
\begin{eqnarray}
N_{\rm ^3He} &=&  \frac{3^{3/2}}{4}\left(\frac{2\pi}{m_0T_\text{eff}}\right)^{3}
N_n\langle n_p\rangle^2 \left(1+\Delta n_p+2\alpha\Delta n\right), \notag \\ \label{he3}  \\
N_{^4\text{He}}
&=&\frac{1}{2}\left(\frac{2\pi}{m_0T_{\rm eff}}\right)^{9/2}N_p\langle n_p\rangle\langle n\rangle^2   \notag \\
&\times& \left[1+(1+4\alpha)\Delta n+\Delta n_p  +\frac{\langle (\delta n\delta n_p)^2\rangle}{\langle n\rangle^2\langle n_p\rangle^2}\right],
\label{he4}
\end{eqnarray}
which further depend on the proton average density $\langle n_p\rangle$, its relative density fluctuation $\Delta n_p=\langle (\delta n_p)^2\rangle /\langle n_p\rangle^2$ and higher-order fluctuations. In Eq.~(\ref{he4}), terms like $\langle (\delta n)^2\delta n_p\rangle$ and $\langle (\delta n_p)^2\delta n\rangle$ are neglected.

Eqs.~(\ref{Eq11})-(\ref{he4}) show that large density fluctuations can affect the yields of light nuclei in relativistic heavy-ion collisions and lead to an $A$ dependence different from $\langle n\rangle^A$ that is expected from the statistical model~\cite{Arm99}. Existing experimental data from the Alternating Gradient Synchrotron (AGS) at $\sqrt{s_{NN}}=4.8$ GeV have shown a striking exponential behavior with a penalty factor of about $50$ per additional nucleon to the produced nuclear cluster up to $A=7$~\cite{Arm99}. Similarly, such a regular exponential behavior is seen at RHIC energies for $A\le 4$~\cite{Sta11}. These results have thus ruled out large nucleon density fluctuations at kinetic freeze-out in heavy-ion collisions at AGS and RHIC top energies.

However, recently published results on light nuclei production in central Pb+Pb collisions at SPS energies~\cite{Ant16} show a quite different behavior. This can be seen from the collision energy dependence of $\mathcal{O}_\text{p-d-t}$ and $\Delta n$. Table~\ref{tab1} summarizes the yields~($dN/dy$ at midrapidity) of $p$, d, $^3$He and $^3$H as well as the yield ratio $^3\text{H}/^3\text{He}$ measured in central Pb+Pb collisions at $20$ AGeV ($0-7\%$ centrality), $30$ AGeV~($0-7\%$ centrality), $40$ AGeV ($0-7\%$ centrality), $80$ AGeV ($0-7\%$ centrality), and $158$ AGeV ($0-12\%$ centrality) by the NA49 Collaboration~\cite{Ant16}. In obtaining the yield of $^3$H, we have used the relation $^3$H=$^3$He$\times$$^3$H/$^3$He. The derived $\mathcal{O}_\text{p-d-t}$ is also shown in Table~\ref{tab1} with errors estimated by assuming they are dominated by correlated systematic errors as a result of similar detector acceptance and phase-space extrapolation.
It is seen from Table~\ref{tab1} that the energy dependence of $\mathcal{O}_\text{p-d-t}$ shows a possible non-monotonic behavior with its largest value at $\sqrt{s_{NN}}=8.8$~GeV. However, it should be pointed out that the evidence for the non-monotonic behavior may not be statistically significant due to the sufficiently large uncertainty. Indeed, the value of $\mathcal{O}_\text{p-d-t}$ at $\sqrt{s_{NN}}=8.8$~GeV deviates by only about $2.5\sigma$ from a $\chi^2$ fit of $\mathcal{O}_\text{p-d-t}$ at $\sqrt{s_{NN}}=$6.3 GeV, 7.6 GeV, 12.3 GeV and 17.3 GeV by the constant $0.471\pm 0.018$.

Equation~(\ref{Eq13}) shows that for a fixed value of $\mathcal{O}_\text{p-d-t}$, the extracted value
for $\Delta n$ depends on the value of $\alpha$.  We note that Eq.~(\ref{Eq13}) has no
solution when $\alpha$ is larger than $\sim 0.23$ at $\sqrt{s_{NN}}=8.8$~GeV.  This feature
suggests that a perfect or strong correlation between neutron and proton density fluctuations
at kinetic freeze-out (i.e., $\alpha = 1$ or $\alpha > 0.23$) cannot appear in collisions at $\sqrt{s_{NN}}=8.8$~GeV. Similar features are also seen at other four collision energies, although the maximum values of $\alpha$ are larger, i.e., $0.32$ for $6.3$~GeV, $0.28$ for $7.6$~GeV, $0.32$ for $12.3$~GeV and $0.29$ for $17.3$~GeV. Table~\ref{tab2} shows the extracted values of $\Delta n$ for $\alpha = -0.2$, $-0.1$, $0$, $0.1$ and $0.2$ at different collisions energies. For all these values of $\alpha$, a similar non-monotonic behavior is seen in the dependence of $\Delta n$ on the collision energy with a peak at $\sqrt{s_{NN}}=8.8$~GeV. Also, the obtained value of $\Delta n$ is much larger than that due to the event-by-event statistical fluctuation in the neutron multiplicity, which is expected to be inversely proportional to its mean value and is thus only about a few per cent.

\begin{figure}[!h]
\includegraphics[scale=0.32]{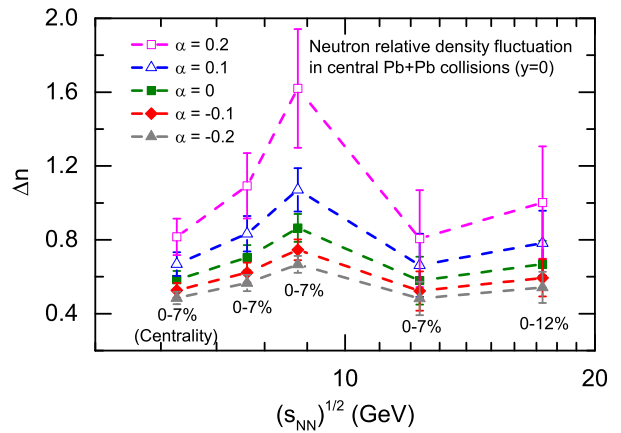}
\caption{Collision energy dependence of the neutron relative density fluctuation $\Delta n$ in central Pb+Pb collisions at SPS energies based on data from Ref.~\cite{Ant16}. Results for $\alpha = -0.2$, $-0.1$, $0$, $0.1$ and $0.2$ are shown by various dotted lines.}
\label{Fig1}
\end{figure}

To see more clearly the collision energy dependence of $\Delta n$, we plot in Fig.~\ref{Fig1} the extracted $\Delta n$ as a function of $\sqrt{s_{NN}}$ for $\alpha = -0.2$, $-0.1$, $0$, $0.1$ and $0.2$. The extracted $\Delta n$ is seen to increase with increasing value of $\alpha$, and the increase is faster for larger value of $\mathcal{O}_\text{p-d-t}$. It is interesting to see that the peak at $\sqrt{s_{NN}}=8.8$ GeV seems to always exist for all values of $\alpha$ considered here. Estimating the statistical significance of the non-monotonic structure of the collision energy dependence of $\Delta n$ by the same method as in the analysis of $\mathcal{O}_\text{p-d-t}$, we find the deviation of the $\Delta n$ value at $\sqrt{s_{NN}}=8.8$~GeV from the average value at the other four energies is about 2.3$\sigma$, 2.5$\sigma$, 2.4$\sigma$, 2.4$\sigma$ and 2.1$\sigma$ for $\alpha = -0.2$, $-0.1$, $0$, $0.1$ and $0.2$, respectively. Given that the present statistical evidence is still weak, it is extremely important to confirm or rule out this possible non-monotonic behavior of the collision energy dependence of $\Delta n$ in future measurements with higher precision.

The possible non-monotonic behavior of the collision energy dependence of $\Delta n$ can be understood as follows. For central Pb+Pb collisions at higher incident energies (e.g., $\sqrt{s_{NN}}=17.3$ GeV and $12.3$ GeV), the reaction system may undergo a crossover rather than a first-order phase transition between the QGP and the hadronic matter, and the density fluctuation in the produced matter is thus insignificant. With decreasing incident energy (e.g., around $\sqrt{s_{NN}}=8.8$ GeV), the reaction system may pass by or approach closely to the CEP and thus develop the largest density fluctuation. With further decrease in the incident energy (e.g., at $\sqrt{s_{NN}}=7.6$ GeV and $6.3$ GeV), the reaction system may move away from the CEP and barely cross the first-order transition line, and the density fluctuation decreases as a result of the smaller size and shorter lifetime of the QGP at lower energies. When the incident energy is further lowered, the reaction system may miss the first-order transition line and no QGP to hadronic matter transition occurs in the collisions, thus resulting in negligible density fluctuation at the kinetic freeze-out. Therefore, the possible non-monotonic behavior shown in Fig.~\ref{Fig1} is consistent with the scenario that the CEP may be reached or closely approached in the produced QGP during its time evolution in central Pb+Pb collisions around $\sqrt{s_{NN}}= 8.8$ GeV.

In the above, we have assumed that there is no energy dependence of $\alpha$ in collisions at SPS energies. In general, the correlation between the neutron and proton density fluctuations, characterized by the value of $\alpha$, near the critical region (e.g., around $\sqrt{s_{NN}}= 8.8$ GeV) is likely larger than those at other collision energies, and thus the extracted $\Delta n$ from Eq.~(\ref{Eq13}) could be larger and the peak structure would become more pronounced. From the parametrization in Ref.~\cite{Cle06} for the chemical freeze-out conditions based on the statistical model fit to available experimental data, the temperature and baryon chemical potential at $\sqrt{s_{NN}}= 8.8$~GeV are estimated to be $T \sim 144$ MeV and $\mu_B \sim 385$ MeV. It is interesting to note that the estimated $\mu_B \sim 385$ MeV for CEP is close to those predicted from the LQCD~\cite{Fod04} and Dyson-Schwinger equation~(DSE)~\cite{Xin14} as well as that based on the hadronic bootstrap approach~\cite{Ant03}. Also, the collision energy $\sqrt{s_{NN}}=8.8~$GeV corresponds to that at which a peak is seen in the measured $K^+/\pi^+$ ratio by the NA49 Collaboration~\cite{Alt08}, which has been interpreted as a signature for the onset of QGP formation~\cite{Gaz99} or the restoration of chiral symmetry~\cite{Cas16} in these collisions.

Although the present study is based on the simple formulas given by Eq.~(\ref{Eq4}) and Eq.~(\ref{Eq5}), the possible non-monotonic behavior in the relative neutron density fluctuation extracted from the measured yield ratio $\mathcal{O}_\text{p-d-t}$ will still be present if the more accurate formula (with $g_\text{rel}$ and $g_\text{size}$~\cite{KJSUN17}) in Eq.~(\ref{Eq1}) is used. This is because the variation in the value of $g$ in Eq.~(\ref{Eq13}) after taking into account the effects due to $g_\text{rel}$ and $g_\text{size}$ is less than $10\%$ for the SPS energies considered here. In addition, although the correlation between neutron and proton density fluctuations influences the value of the extracted $\Delta n$, it does not change the non-monotonic behavior of $\Delta n$ as a function of the collision energy. Our study is, however, based on one set of experimental data with large uncertainties and a simplified model. Further experimental and theoretical investigations are needed to verify the present results and eventually establish the yield ratio $\mathcal{O}_\text{p-d-t} = N_{^3\text{H}} N_p / N_\text{d}^2$ as a robust probe to the QCD critical endpoint. These include the experimental BES program at RHIC in the energy range considered here with high luminosity beams as well as detectors of excellent particle identification and large acceptance, and theoretical modeling of light nuclei production and its connection to baryon density fluctuations.

In summary, with a newly derived analytical coalescence formula for cluster production in heavy-ion collisions, we have demonstrated that information on the relative density fluctuation of neutrons~($\Delta n= \langle (\delta n)^2\rangle/\langle n\rangle^2$) at kinetic freeze-out can be determined directly from the yield ratio $\mathcal{O}_\text{p-d-t} = N_{^3\text{H}} N_p / N_\text{d}^2$. From measured yields of light nuclei at SPS energies by the NA49 Collaboration, we have extracted the collision energy dependence of $\Delta n$ and found a possible non-monotonic behavior with a peak at $\sqrt{s_{NN}}= 8.8$~GeV, suggesting that the CEP in the QCD phase diagram may have been reached or closely approached in these collisions with its temperature and baryon chemical potential estimated to be $T^\text{CEP} \sim 144$ MeV and $\mu^\text{CEP}_B \sim 385$ MeV, respectively. Given that the present statistical evidence for the peak structure of the collision energy dependence of $\Delta n$ is still weak, future measurements of light nuclei production in the BES program at RHIC are extremely useful to confirm the present observations and to more precisely determine the location of the CEP in the QCD phase diagram.

\begin{acknowledgments}

The authors thank Vadim Kolesnikov and Peter Seyboth for providing the experimental data. This work was supported in part by the Major State Basic Research Development Program (973 Program) in China under Contract Nos. 2015CB856904 and 2013CB834405, the National Natural Science Foundation of China under Grant Nos. 11625521, 11275125 and 11135011, the Program for Professor of Special Appointment (Eastern Scholar) at Shanghai Institutions of Higher Learning, Key Laboratory for Particle Physics, Astrophysics and Cosmology, Ministry of Education, China, the Science and Technology Commission of Shanghai Municipality (11DZ2260700), the US Department of Energy under Contract No. DE-SC0015266 and No. DE-SC0012704, as well as the Welch Foundation under Grant No. A-1358 and Shandong University.

\end{acknowledgments}

\end{document}